\documentclass{egpubl}
\usepackage{eg2025}
 
\usepackage[T1]{fontenc}
\usepackage{dfadobe}  

\usepackage{cite}
\BibtexOrBiblatex
\electronicVersion
\PrintedOrElectronic
\ifpdf \usepackage[pdftex]{graphicx} \pdfcompresslevel=9
\else \usepackage[dvips]{graphicx} \fi

\usepackage{egweblnk} 

\usepackage{mathtools,stmaryrd,nicefrac}
\usepackage[many]{tcolorbox}
\usepackage{ifthen,comment,xspace}
\usepackage{overpic}
\usepackage{subcaption}
\usepackage{multirow}

\usepackage[outline]{contour}
\contourlength{0.075em}

\usepackage[linesnumbered,ruled,vlined]{algorithm2e} %

\SetAlFnt{\small}
\SetAlCapFnt{\small}
\SetAlCapNameFnt{\small}
\SetAlCapHSkip{0pt}
\let\oldnl\nl%
\newcommand{\nonl}{\renewcommand{\nl}{\let\nl\oldnl}}%
\definecolor{cmtClr}{HTML}{028A0F}

\usepackage{tikz}
\usetikzlibrary{calc}

\usepackage{enumitem}
\setlist[itemize]{parsep=0pt,partopsep=0pt,leftmargin=*,itemsep=5pt}
\setlist[enumerate]{parsep=0pt,partopsep=0pt,leftmargin=*,itemsep=5pt}

\usepackage{booktabs} 
\usepackage{amsmath}
\usepackage{amssymb}

\newif\ifpaperdraft
\paperdraftfalse

\title{A Survey on Physics-based Differentiable Rendering}

\author[Y. Zeng, G. Cai \& S. Zhao]
{\parbox{\textwidth}{\centering 
    Y. Zeng$^{1}$,
    G. Cai$^{2}$
    and S. Zhao$^{2}$
}
\\
\parbox{\textwidth}{\centering 
    $^1$The Hong Kong University of Science and Technology, China\\
    $^2$University of California Irvine, United States of America
}
}

\definecolor{orange_cubic}{rgb}{.9765, .5887, .3569}
\definecolor{purple_cubic}{rgb}{.4706, 0, .5216}
\definecolor{green_cubic}{rgb}{.28603, .81178, .5008}

\definecolor{grayLL}{rgb}{.98, .98, .98}
\definecolor{grayL}{rgb}{.9, .9, .9}
\definecolor{purpleL}{rgb}{.9735, .95, .9761}
\definecolor{purpleD}{rgb}{.8941, .8, .9043}
\definecolor{greenL}{rgb}{.9643, .9906, .9750}
\definecolor{greenD}{rgb}{.7145, .9249, .7999}
\definecolor{greenDD}{rgb}{.3145, .6249, .3999}
\definecolor{orangeLL}{rgb}{0.9991, 0.9846, 0.9759}
\definecolor{orangeL}{rgb}{.9982, .9692, .9518}
\definecolor{orangeD}{rgb}{.9929, .8766, .8071}

\definecolor{redL}{rgb}{1.0, 0.95, 0.95}
\definecolor{redD}{rgb}{1.0, 0.8, 0.8}
\definecolor{redDD}{rgb}{1.0, 0.4, 0.4}
\definecolor{yellowL}{rgb}{1.0, 1.0, 0.95}
\definecolor{yellowD}{rgb}{0.95, 0.95, 0.6}
\definecolor{yellowDD}{rgb}{0.8, 0.8, 0.2}
\definecolor{blueLL}{rgb}{0.98, 0.98, 1.0}
\definecolor{blueL}{rgb}{0.95, 0.95, 1.0}
\definecolor{blueD}{rgb}{0.8, 0.8, 1.0}
\definecolor{blueDD}{rgb}{0.6, 0.6, 1.0}

\definecolor{uciBlue}{RGB}{0,100,164}
\definecolor{uciBlueL}{RGB}{127.5, 177.5, 209.5}
\definecolor{uciOrange}{RGB}{247,141,45}
\definecolor{uciOrangeL}{RGB}{251,198,150}

\newtcolorbox{myBox}{%
	colback=grayLL,colframe=grayL,top=1mm,bottom=1mm,left=1mm,right=1mm%
}
\newtcolorbox{myTitledBox}[2][]{%
	colback=grayLL,colframe=grayL,top=1mm,bottom=1mm,left=1mm,right=1mm,enlarge top by=0.5em,title={#2},fonttitle=\bfseries\small\color{gray},#1%
}

\tcolorboxenvironment{myitemizeP}{blanker,before skip=6pt,after skip=6pt,borderline west={3mm}{0pt}{purpleD}}

\tcolorboxenvironment{myitemizeG}{blanker,before skip=6pt,after skip=6pt,borderline west={3mm}{0pt}{greenD}}

\tcolorboxenvironment{myitemizeO}{blanker,before skip=6pt,after skip=6pt,borderline west={3mm}{0pt}{orangeD}}
\newtcbtheorem{thm}{Theorem}%
{colback=white,colframe=orangeD,top=1mm,bottom=1mm,left=1mm,right=1mm,enlarge top by=1mm,fonttitle=\bfseries\color{black}}{thm}
\newcommand{\mymathbox}[3]{%
    \tcboxmath[top=0mm,bottom=0mm,left=0mm,right=0mm,fonttitle=\bfseries\scriptsize\color{black},colbacktitle=white,enhanced,attach boxed title to top center={yshift=-1mm},boxed title style={top=0mm,bottom=0mm,left=0mm,right=0mm},colframe=#1,colback=white,title=#2]{#3}
}

\renewcommand{\to}{\!\shortrightarrow\!}

\newcommand{\defeq}{\vcentcolon=}
\newcommand{\eqdef}{=\vcentcolon}
\makeatletter
\newcommand{\dotr}[1]{%
	\mathpalette\@dotr{#1}%
}
\newcommand*{\@dotr}[2]{%
	\sbox0{$\m@th#1#2$}%
	\usebox{0}%
	\raisebox{\dimexpr\ht0-\height}{$\m@th#1\@smallbullet#1\bullet$}%
	\kern\scriptspace
}
\newcommand*{\@smallbullet}[2]{%
	\scalebox{.4}{$\m@th#1#2$}%
}
\makeatother

\newcommand{\ttx}{\mathtt{X}}

\newcommand{\D}{\mathrm{d}}

\newcommand{\f}{f}

\newcommand{\Le}{L_\mathrm{e}}

\newcommand{\We}{W_\mathrm{e}}
\newcommand{\bsdf}{f_\mathrm{s}}

\newcommand{\pdf}{\mathrm{pdf}}

\newcommand{\intbox}[1]{\mymathbox{uciBlueL}{interior}{#1}}
\newcommand{\bndbox}[1]{\mymathbox{uciOrangeL}{boundary}{#1}}

\newlength{\tikzLen}

\newlength{\resLen}

\newcommand{\omegai}{\omega_\mathrm{i}}
\newcommand{\omegao}{\omega_\mathrm{o}}
\newcommand{\omegah}{\omega_\mathrm{h}}
\newcommand{\omegab}{\omega_\mathrm{b}}
\newcommand{\omegap}{\omega_\mathrm{p}}
\newcommand{\omegabnd}{\omega^\mathrm{B}}
\newcommand{\fv}{f_\mathrm{v}}
\newcommand{\fs}{f_\mathrm{s}}
\newcommand{\fp}{f_\mathrm{p}}
\newcommand{\hatfv}{\hat{f}_\mathrm{v}}
\newcommand{\pcam}{p_\mathrm{cam}}
\newcommand{\ncam}{n_\mathrm{cam}}
\newcommand{\Lo}{L_\mathrm{o}}
\newcommand{\Li}{L_\mathrm{i}}
\newcommand{\sigmas}{\sigma_\mathrm{s}}
\newcommand{\vp}{v_\mathrm{p}}
\newcommand{\paths}{\bar{p}^\mathrm{S}}
\newcommand{\pathd}{\bar{p}^\mathrm{D}}
\newcommand{\ps}{p^\mathrm{S}}
\newcommand{\pd}{p^\mathrm{D}}
\newcommand{\xs}{x^\mathrm{S}}
\newcommand{\xd}{x^\mathrm{D}}
\newcommand{\xb}{x^\mathrm{B}}

\ifpaperdraft
	
\else
	
\fi

\begin{document}
\maketitle

\begin{abstract}

Physics-based differentiable rendering has emerged as a powerful technique in computer graphics and vision, with a broad range of applications in solving inverse rendering tasks. At its core, differentiable rendering enables the computation of gradients with respect to scene parameters, allowing optimization-based approaches to solve various problems. Over the past few years, significant advancements have been made in both the underlying theory and the practical implementations of differentiable rendering algorithms. In this report, we provide a comprehensive overview of the current state of the art in physics-based differentiable rendering, focusing on recent advances in general differentiable rendering theory, Monte Carlo sampling strategy, and computational efficiency. 


\end{abstract}

\section{Introduction}
\label{sec:intro}

In recent years, physics-based differentiable rendering (PBDR) has received significant attention in both the computer graphics and vision communities. By enabling the computation of gradients with respect to scene parameters, differentiable rendering provides a powerful tool for solving inverse rendering problems. 
At its core, differentiable rendering extends modern physics-based rendering techniques by allowing gradients to be computed through the light transport simulation process, typically by leveraging Monte Carlo methods to handle the integrals that are unique to differentiable rendering. These advances have unlocked new capabilities in optimization and learning-driven tasks, making PBDR a powerful tool in many graphics and vision workflows.

At the core of many differentiable systems is automatic differentiation (auto-diff), which propagates gradients efficiently through computational graphs by applying the chain rule. While auto-diff is highly effective for simpler differentiable rendering tasks, it encounters significant challenges when applied to complex light transport scenarios. Modern physics-based rendering that leverages Monte Carlo integration often involves phenomena that are not trivially differentiable such as occlusion, geometric boundary, and participating media, which introduce discontinuities and bias in the auto-diff process. Solving these problems and achieving an unbiased estimation of gradient has been one of the main goals of differentiable rendering.

Over the past few years, significant progress has been made in the underlying theory to address these challenges. To directly \textit{analyze} the discontinuities, Li et al.~\cite{Li:2018:DMC} introduce Monte Carlo edge sampling, achieving unbiased differentiation of the rendering equation~\cite{Kajiya:1986:RE} for the first time. Zhang et al.~\cite{Zhang:2019:DTRT} extend this method to differentiate the radiative transfer equation, enabling volumetric lighting effects. Zhang et al.~\cite{Zhang:2020:PSDR} then propose a more general framework known as path-space differentiable rendering (PSDR), which differentiates the path integral formulation~\cite{veach1997robust} and introduces methods to track discontinuities in light paths, significantly improving performance and enabling the development of more specialized Monte Carlo estimators. This framework is also extended to support more scenarios such as participating media \cite{Zhang:2021:PSDR} and implicit surfaces \cite{Zhou:2024:PSDR-SDF}. PSDR for time-gated rendering is also introduced in \cite{Wu:2021:Time, Yi:2021:Transient}, but we are mainly focusing on steady-state scenarios and will not thoroughly discuss these works in this survey. 

Concurrently, another route that tries to \textit{avoid} explicit sampling of the discontinuities is first introduced by Loubet et al.~\cite{Loubet:2019:Reparameterizing}, using an approximated reparameterization technique. This is later refined by Bangaru et al.~\cite{bangaru:2020:warpedsampling} into an unbiased estimation using warped-area sampling. Xu et al.~\cite{Xu:2023:PSDR-WAS} further improve this approach by integrating path sampling into warped-area reparameterization, enabling the use of more advanced Monte Carlo algorithms. Reparameterization techniques are also extended to support signed-distance functions (SDFs) \cite{Vicini:2022:DiffSDF, Bangaru:2022:DiffNeuralSDF}.

Apart from the advancements in the general theory of physics-based differentiable rendering, there also exist various works that focus on improving the efficiency and reducing the variance of the Monte Carlo estimators. We roughly classify them into handling the interior integral \cite{Zhang:2021:Antithetic, Yu:2022:Pixel, Belhe:2024:BRDF, Zeltner:2021:Light, Nimier:2022:Volume} and the boundary integral \cite{Yan:2022:Guiding, Zhang:2023:Projective}, which are two fundamental terms in differentiable rendering that we will discuss in the following sections. In addition, there are also works that optimize the gradient backpropagation process for differentiable rendering algorithms \cite{Nimier:2020:Radiative, Vicini:2021:PathReplay}. 

Our survey is organized as follows. In Section~\ref{sec:theory}, we first give a detailed introduction to the general theories of physics-based differentiable rendering. In Section~\ref{sec:interior}, we present several methods that focus on efficient Monte Carlo estimation of the interior term and a few optimization strategies on gradient propagation. In Section~\ref{sec:boundary}, we introduce a few studies on custom sampling strategies for the boundary term. Finally, in Section \ref{sec:conclusion}, we draw conclusions and suggest possible research directions in the future.

\section{General theory of physics-based differentiable rendering}
\label{sec:theory}

\subsection{Differential spherical integral}
\label{sec:theory_spherical}

Physics-based rendering simulates the interaction of light with objects in a scene to generate photorealistic images. The fundamental principle of this technology is rooted in the physics of light transport. One common approach to describe this principle is the rendering equation~\cite{Kajiya:1986:RE}, which quantifies the radiance exiting in a scene as a function of source emission, incoming radiance, and surface properties. The rendering equation can be expressed as:
\begin{equation}
\label{eq:RE}
    \Lo(x, \omegao) = \Le(x, \omegao) + \int_{\mathbb{S}^2}  \Li(x, \omegai) \fs(x, \omegai, \omegao) \D \sigma(\omegai),
\end{equation}
where \(\Lo\) is the full exitant radiance at point \( x \) with direction \( \omegao \), \(\Le\) is the self-emitted radiance, \( \mathbb{S}^2 \) is the surface of a unit sphere
, \(\Li\) is the incident radiance, \(\fs\) is the cosine-weighted bidirectional scattering distribution function (BSDF) and \( \D\sigma \) is the solid-angle measure. 

For physics-based \textit{differentiable} rendering, the primal goal is to differentiate the rendering equation w.r.t. an arbitrary scene parameter \( \pi \):
\begin{equation}
\label{eq:diff_RE}
    \frac{\partial \Lo}{\partial \pi} = \\ \frac{\partial \Le}{\partial \pi} + \frac{\partial}{\partial \pi} \int_{\mathbb{S}^2}  \Li \fs \D \sigma(\omegai),
\end{equation}
where \( \Le, \Li \) and \( \fs \) might be a function of \( \pi \). For instance, when parameterizing light source position or other geometry-related parameters with \( \pi \), \( \Li \) often becomes dependent on \( \pi \). When parameterizing BSDF related parameters such as albedo with \( \pi \), \( \fs \) becomes dependent on \( \pi \). 
Note that \( \Le \) is a term related to the surface itself, and its derivative has to be computed specifically according to the surface properties. Fortunately, \( \partial \Le / \partial \pi \) can be trivially obtained by auto-diff in most cases. Therefore, the key objective of differentiable rendering comes down to estimating the derivative of this \textit{spherical integral}:
\begin{equation}
\label{eq:diff_RE_int}
    \frac{\partial I}{\partial \pi} = \frac{\partial}{\partial \pi} \int_{\mathbb{S}^2}  \Li \fs \D \sigma(\omegai).
\end{equation}
Unfortunately, this integral cannot be trivially differentiated, due to the discontinuity in the integrand. One common case of discontinuity lies in the incident radiance \( \Li \) that comes from different directions. For example, when a part of a light source is occluded by another object that is evolving with \( \pi \), the silhouette of the object becomes a boundary that causes discontinuity, where the incident radiance coming from different sides of the silhouette is discontinuous. Since the silhouette is moving with \( \pi \), it contributes to the derivative in Eq.~\ref{eq:diff_RE_int}. This type of discontinuity is also called \textit{visibility discontinuity}, and the silhouette of such objects is called \textit{visibility boundary}. 

\setlength{\resLen}{7in}
\begin{figure}[t]
    \centering
    \small
    \addtolength{\tabcolsep}{-3pt}
    \begin{tabular}{c}
         \includegraphics[trim={0 30 0 0},clip]{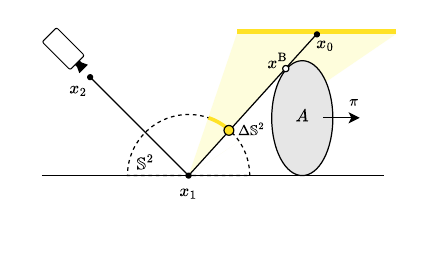}
    \end{tabular}
    
    \caption{\label{fig:spherical_integral}
        \textbf{Differential spherical integral:} in this simple 2D scene, object A is moving with \( \pi \), occluding part of the yellow light source. Considering \( x_1 \) as the shading point, \( \xb \) becomes a discontinuity boundary, and its projection on the unit sphere \( \mathbb{S}^2 \) becomes part of \( \Delta\mathbb{S}^2 \). The gradient caused by the motion of \( \Delta\mathbb{S}^2 \) can not be estimated by the interior integral, and needs to be treated separately in the boundary integral. 
    }
\end{figure}

To correctly compute the differentiation in Eq.~\ref{eq:diff_RE_int}, one intuitive way is to directly identify the boundaries that cause discontinuity in the scene and estimate their contribution to the differentiation. This idea is first introduced by Li et al.~\cite{Li:2018:DMC}, and later formalized by Zhang et al.~\cite{Zhang:2019:DTRT} using Reynolds Transport Theorem. The idea is to decompose the differentiation into two parts:
\begin{equation}
\label{eq:diff_RE_Reynolds}
    \frac{\partial I}{\partial \pi} = \intbox{\int_{\mathbb{S}^2} \frac{\partial (\Li \fs)}{\partial \pi}\D\sigma(\omegai)} + \bndbox{\int_{\Delta\mathbb{S}^2} \Delta (\Li \fs) V \D \ell(\omegai) },
\end{equation}
where \( \Delta\mathbb{S}^2 \) is the union of all discontinuity boundaries projected to \( \mathbb{S}^2 \), \( 
\Delta f = f^- - f^+ \) denotes the difference of function \( f \) when approaching the boundary from two sides, \( V = n^\partial \cdot v^\partial \) is the scalar normal velocity, or the value of boundary's velocity \( v^\partial \) (evolving with \( \pi \)) along the boundary's normal direction \( n^\partial \), and \( \D \ell \) is the curve-length measure. Figure~\ref{fig:spherical_integral} illustrates a simple scene that involves such boundaries.

Through this decomposition, the interior integral can be estimated using forward rendering strategies combined with auto-diff. On the other hand, estimating the boundary integral would require finding the discontinuity boundaries first, before estimating their contribution. Li et al.~\cite{Li:2018:DMC} propose the first approach to estimate the boundary integral by directly sampling the boundary. For scenes represented with meshes, one key observation is that discontinuities always occur at triangle edges. If the shading is smooth, it narrows further down to the silhouette edges. Therefore, a straightforward approach to estimate the boundary integral would be to explicitly sample the edges in the scene, and this strategy is known as \textit{edge sampling}. 

For primary boundary that is directly observed by the detector, a simple way is to project and clip all meshes on screen space, and sample from the visible edges. However, for secondary boundary that occurs in light bounces, since the silhouette of an object is dependent on the shading point, it would be inefficient to do the projection for every light bounce. To address this issue, Li et al. introduce an importance sampling scheme which uses a hierarchy to quickly reject interior edges and find the silhouette, drawing inspiration from \cite{Sander:2000:Silhouette}. 

In addition to handling light transport on mesh surfaces, Zhang et al.~\cite{Zhang:2019:DTRT} extend edge sampling to support participating media. This is done by differentiating the radiative transfer equation,
\begin{equation}
\label{eq:radiative}
L_o = \mathcal{K}_\mathrm{T} \mathcal{K}_\mathrm{C} \Li + Q,
\end{equation}
where \( \mathcal{K}_\mathrm{T} \) is the transport operator, \( \mathcal{K}_\mathrm{C} \) is the collision operator, and \( Q \) is the source term. 

After deriving the derivatives of each individual term, Zhang et al. propose the complete equation that differentiates Eq.~\ref{eq:radiative}. Due to its complexity, we refer readers to Eq. 32 of the original paper for the full equation. The key idea of this differentiation is still decomposing the differentiation into an interior and a boundary term. To estimate this differentiation, Zhang et al. combine the edge sampling process with volumetric path tracing that is commonly used in modern forward rendering.

\subsection{Differential path integral}
\label{sec:theory_psdr}

Starting from the rendering equation, the theory of differential spherical integral and the edge sampling strategy seems to be an intuitive and practical solution to differentiate rendering. However, since the rendering equation only describes a single light scatter event within a scene, the estimation of its derivative using Eq.~\ref{eq:diff_RE_Reynolds} and edge sampling needs to be performed for every light bounce, in order to achieve an unbiased result. This will inevitably cause performance issues when scaling to multi-bounce lighting and dense meshes.

To this end, Zhang et al.~\cite{Zhang:2020:PSDR} propose path-space differentiable rendering (PSDR), a new method to estimate the derivative in a decoupled manner that does not require boundary integral estimation for every light bounce. PSDR starts from differentiating the \textit{path integral} formulation \cite{veach1997robust}, which depicts the response \( I \) of a radiometric detector on the domain of the whole scene:
\begin{equation}
\label{eq:path_int}
I = \int_\Omega f(\Bar{x}) \D \mu(\Bar{x}).
\end{equation}
In this equation,
\begin{itemize}
    \item \( \Bar{x} \defeq (x_0, x_1, \dots, x_N) \) is a light path containing \( N + 1 \) vertices on object surfaces, with \( x_0 \) on a light source and \( x_N \) on the detector;
    \item \( \Omega \defeq \bigcup_{N=1}^\infty \mathcal{M}^{N+1} \) is the path space containing all possible light paths of finite lengths within a scene, where \( \mathcal{M} \) is the union of all object surfaces;
    \item \( f(\Bar{x}) \) is the measurement contribution function; 
    \item \( \D \mu(\Bar{x}) \defeq \prod_{n=0}^N \D A(x_n) \) is the area-product measure, where \( A \) is the surface area measure.
\end{itemize}
The measurement contribution \( f(\Bar{x}) \) can be expressed as
\begin{equation}
\label{eq:path_measurement}
f(\Bar{x}) \defeq \left[ \prod_{n=0}^N \fv(\Bar{x}, n) \right] \left[ \prod_{n=1}^N G(x_{n-1} \leftrightarrow x_n) \right].
\end{equation}
In this equation,  
\begin{equation}
\label{eq:path_measurement_radiance_surface}
\fv(\Bar{x}, n) \defeq 
\begin{cases}
\fs(x_{n-1} \rightarrow x_n \rightarrow x_{n+1}), & 0 < n < N \\
\Le(x_0 \rightarrow x_1), & n = 0 \\
\We(x_{n-1} \rightarrow x_n), & n = N
\end{cases}
\end{equation}
where \( \fs \) is the bidirectional scattering distribution function (BSDF), \( \Le \) is the source emission, and \( \We \) is the sensor importance. \( G(x_{n-1} \leftrightarrow x_n) \) is the geometric term expressed as
\begin{equation}
\label{eq:path_measurement_geometry_surface}
G(x \leftrightarrow y) \defeq \mathbb{V} (x \leftrightarrow y) \frac{|n(x) \cdot \overrightarrow{xy}||n(y) \cdot \overrightarrow{xy}|}{\lVert x - y \rVert ^2},
\end{equation}
where \( \mathbb{V} \) is the mutual visibility function, \( n(x) \) is the surface normal at \( x \), and \( \overrightarrow{xy} \defeq (y - x) / \lVert y - x \rVert \) is the normalized direction from \( x \) to \( y \).

\setlength{\resLen}{3.6in}
\begin{figure}[t]
    \centering
    \small
    \addtolength{\tabcolsep}{-3pt}
    \begin{tabular}{c}
         \includegraphics[trim={20 10 0 0},clip, width=\resLen]{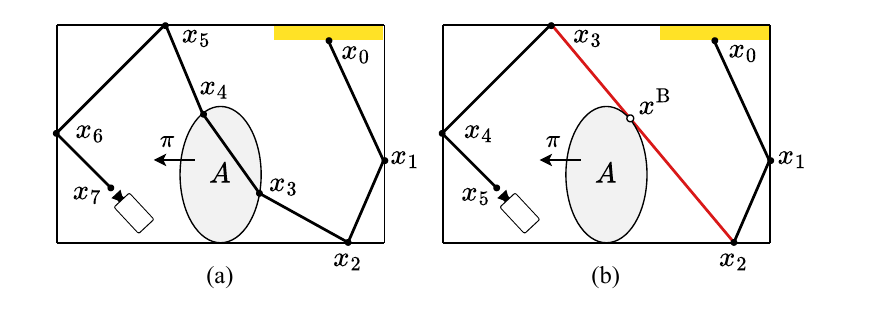}
    \end{tabular}
    
    \caption{\label{fig:psdr}
        \textbf{Differential path integral:} in this simple 2D scene, object A is moving with \( \pi \). (a) The light path is an ordinary light path that does not involve a boundary segment. (b) The light path is a boundary light path containing a boundary segment \( (x_2, x_3) \), and therefore the material form of the light path belongs to \( \partial \hat{\Omega} \).
    }
\end{figure}

PSDR aims to decompose the differentiation of this integral into an interior and a boundary term, similar to edge sampling, but on a global domain. To achieve this, PSDR proposes material-form parameterization. This parameterization introduces a union of reference surfaces \( \mathcal{B} \) that is independent of \( \pi \), and uses a differentiable one-to-one mapping function \( \ttx(\cdot, \pi) \) that maps any point \( p \) on \( \mathcal{B} \) to a point \( x \) on the original object surfaces \( \mathcal{M} \) that might be evolving with \( \pi \). This mapping also ensures that when \( \pi=\pi_0 \), which is the current value in the primal scene, the mapping is an identity map. In this way, the primal scene is not modified by the mapping, but its motion is changed. In practice, we simply use an identity map for the motion, but the Jacobians for this reparameterization still need to be computed correctly. 
With this mapping, the path integral can be parameterized to its material form:
\begin{equation}
\label{eq:path_int_material}
I = \int_{\hat{\Omega}} \hat{f}(\Bar{p}) \D \mu(\Bar{p}).
\end{equation}
In this equation,
\begin{itemize}
    \item \( \Bar{p} \defeq (p_0, p_1, \dots, p_N) \) is a material light path containing \( N + 1 \) vertices on \( \mathcal{B} \);
    \item \( \hat{\Omega} \defeq \bigcup_{N=1}^\infty \mathcal{B}^{N+1} \) is the material path space;
    \item \( \hat{f}(\Bar{p}) \) is the material measurement contribution function, which can be expressed as
    \begin{equation}
    \label{eq:psdr_material_measurement_contribution}
    \hat{f}(\Bar{p}) \defeq \left[ \prod_{n=0}^N \hatfv(\Bar{p}, n) \right] \left[ \prod_{n=1}^N G(x_{n-1} \leftrightarrow x_n) \right],
    \end{equation}
    where \( \hatfv(\Bar{p}, n) \defeq \fv(\bar{x}, n) J(p_n) \) and \( J(p_n) \defeq \lVert \D A(x_n) / \D A(p_n) \rVert \).
\end{itemize}

This material-form parameterization ensures the domain of integration \( \hat{\Omega} \) is independent of \( \pi \), and the geometric boundaries of the surfaces in \( \mathcal{M} \) no longer contribute to the boundary integral. This means we now have less types of discontinuities to keep track of, and enables the decomposed differentiation using Reynolds transport theorem:
\begin{equation}
\label{eq:diff_path_int}
\frac{\partial I}{\partial \pi} = \intbox{\int_{\hat{\Omega}} \frac{\partial \hat{f}(\Bar{p})}{\partial \pi} \D\mu(\Bar{p})} + \bndbox{\int_{\partial \hat{\Omega}} \Delta \hat{f_K}(\Bar{p}) V_K(p_K) \D\mu'(\Bar{p})}.
\end{equation}
In this equation,
\begin{itemize}
    \item \( \partial \hat{\Omega} \) is the material boundary path space, containing all material light paths \( \Bar{p} \) that involve a boundary segment \( (p_{K-1}, p_K) \) (i.e., a segment that resides on a visibility boundary), which is the cause of the discontinuity in \( \hat{f}(\Bar{p}) \). Figure~\ref{fig:psdr} illustrates a simple scene containing interior and boundary light paths;
    \item \( \Delta \hat{f_k}(\Bar{p}) \defeq \hat{f}(\Bar{p}) \Delta G(x_{K-1} \leftrightarrow x_K) / G(x_{K-1} \leftrightarrow x_K) \) is the material boundary contribution, under the assumption that \( \fv(\hat{p}, n) \) is continuous. In this equation, \( \Delta G \defeq G^- - G^+ \) denotes the difference of \( G \) when approaching the visibility boundary from two directions (along the boundary's normal or its opposite). \( \Delta G = -G \) when the normal points towards the region visible to \( p_{K-1} \), and \( \Delta G = G \) if otherwise;
    \item \( V_K(p_K) = n^\partial \cdot v^\partial \) is the scalar normal velocity of the discontinuity point \( p_K \), or the value of boundary's velocity \( v^\partial \) along the boundary's normal direction \( n^\partial \) at this point.
    \item \( \D\mu'(\Bar{p}) \defeq \D \ell (p_K) \prod_{n \neq K} \D A(p_n) \) is the measure associated with \( \partial \hat{\Omega} \).
\end{itemize}

The equation has a similar structure to edge sampling, but since the integrals involve all material-form light paths within a scene, the estimation of interior integral and boundary integral can now be completely decoupled. This offers a few key benefits: (a) The interior integral can be estimated using conventional path sampling methods (e.g., unidirectional path-tracing), making it possible to be computed jointly in a forward rendering pass. (b) The boundary integral can be estimated using a specialized Monte Carlo estimator, which helps to reduce variance and improve performance. (c) The number of samples used to estimate each integral can be adjusted according to the scene, achieving a balance between gradient accuracy and performance. 

Evaluating the boundary integral requires sampling boundary segments, which is previously done by computationally expensive silhouette edge detection. To this end, PSDR rewrites the boundary integral to a multi-directional form, and proposes a Monte Carlo estimator that samples boundary light paths in a multi-directional manner. A boundary light path \( \bar{p} \) can be written as \( \bar{p} = (\ps_s, \dots, \ps_0, \pd_0, \dots, \pd_t) \) where \( (\ps_0, \pd_0) \) is the boundary segment, \( \paths \defeq (\ps_s, \dots, \ps_1) \) is the source sub-path, and \( \pathd \defeq (\pd_1, \dots, \ps_t) \) is the detector sub-path. Therefore, the boundary integral can be rewritten into a multi-directional form:
\begin{equation}
\label{eq:psdr_rewrite_bsd_int}
\int_\mathcal{B} \int_{\Delta \mathcal{B}} \bigg[ \underbrace{\int_{\hat{\Omega}} \hat{f}^\mathrm{S} \D\mu(\paths)}_{\eqdef I^\mathrm{S}} \bigg] \hat{f}^\mathrm{B} \bigg[ \underbrace{\int_{\hat{\Omega}} \hat{f}^\mathrm{D} \D\mu(\pathd)}_{\eqdef I^\mathrm{D}} \bigg] \D\ell(\pd_0) \D A(\ps_0),
\end{equation}
with
\begin{alignat}{2}
\label{eq:psdr_rewrite_bsd}
& \hat{f}^\mathrm{B} \; \defeq \;
&& \Delta G(\xs_0 \leftrightarrow \xd_0) V_{\Delta \mathcal{B}}, \\
& \notag \hat{f}^\mathrm{S} \; \defeq \;
&& \hatfv(\ps_1 \rightarrow \ps_0 \rightarrow \pd_0) \\
& && \quad \prod_{n=1}^s \hatfv (\ps_{n+1} \rightarrow \ps_n \rightarrow \ps_{n-1}) G(\xs_{n-1} \leftrightarrow \xs_n), \\
& \notag \hat{f}^\mathrm{D} \; \defeq \;
&& \hatfv(\ps_0 \rightarrow \pd_0 \rightarrow \pd_1) \\
& && \quad \prod_{n=1}^t \hatfv (\pd_{n+1} \rightarrow \pd_n \rightarrow \pd_{n-1}) G(\xd_{n-1} \leftrightarrow \xd_n),
\end{alignat}
where \( \Delta \mathcal{B} \) is the union of visibility boundaries when viewed from \( \ps_0 \), and \( V_{\Delta \mathcal{B}} \) is the scalar normal velocity of \( \pd_0 \) with \( \ps_0 \) fixed. To sample the boundary segment \( (\ps_0, \pd_0) \), we can further perform a change of variable from \( \ps_0 \) and \( \pd_0 \) to \( \xb \) (the intersection point of the boundary segment and the object causing this boundary) and \( \omegabnd \) (the unit direction of \( \ps_0 \rightarrow \pd_0 \)). Then the boundary integral can be expressed as
\begin{equation}
\label{eq:psdr_rewrite_xw}
\int_{\mathcal{E}} \int_{\mathbb{S}^2} I^\mathrm{S} \hat{f}^\mathrm{B} J^\mathrm{B} I^\mathrm{D} \D\sigma(\omegabnd) \D\ell(\xb),
\end{equation}
where \( \mathcal{E} \) is the union of all mesh edges, and \( J^\mathrm{B} \) is the Jacobian determinant of this change of variable. This form of boundary integral essentially implies that we no longer need to manually find the boundary edges during ray tracing. Instead, we first sample an edge and use it as a boundary edge to construct the boundary light path. Therefore, we can now design a Monte Carlo estimator that samples boundary light paths in a multi-directional manner, consisting of three steps: (i) sample an edge within the scene, (ii) generate the boundary segment \( (\ps_0, \pd_0) \) from the edge, and (iii) trace the path in both directions to reach the light source and sensor respectively, completing a boundary light path. More specialized sampling strategies tailored for this process will be discussed in \S\ref{sec:boundary}.

\subsubsection{PSDR of participating media}

While the original PSDR is limited to surface-only light transport, Zhang et al.~\cite{Zhang:2021:PSDR} propose a more generalized version of PSDR to support participating media. This starts from modifying path-integral formulation in Eq.~\ref{eq:path_int}, by slightly changing the definition of \( \Omega \) and \( \D\mu(\Bar{x}) \), as well as modifying \( \fv(\Bar{x}, n) \) and \( G(x_{n-1} \leftrightarrow x_n) \) in the measurement contribution \( f(\Bar{x}) \) to support volumetric light transport:
\begin{multline}
\label{eq:path_measurement_radiance_volume}
\fv(\Bar{x}, n) \defeq \\
\begin{cases}
\fs(x_{n-1} \rightarrow x_n \rightarrow x_{n+1}), & 0 < n < N \;\text{and}\; x_n \in \mathcal{M} \\
\sigmas (x_n) \fp(x_{n-1} \rightarrow x_n \rightarrow x_{n+1}), & 0 < n < N \;\text{and}\; x_n \in \mathcal{V} \\
\Le(x_0 \rightarrow x_1), & n = 0 \\
\We(x_{N-1} \rightarrow x_N), & n = N
\end{cases}
\end{multline}
where \( \mathcal{V} \) is the union of all object's interior volumes, \( \sigmas \) is the scattering coefficient, and \( \fp \) is the single-scattering phase function, and 
\begin{equation}
\label{eq:path_measurement_geometry_volume}
G(x \leftrightarrow y) \defeq \mathbb{V}(x \leftrightarrow y) \tau (x \leftrightarrow y) \frac{D_x(y)D_y(x)}{\lVert x - y \rVert ^2},
\end{equation}
where \( \tau \) is the transmittance function, and
\begin{equation}
\label{eq:path_measurement_geometry_D_volume}
D_x(y) \defeq 
\begin{cases}
    |n(x) \cdot \overrightarrow{xy}|, & x \in \mathcal{M} \\
    1. & x \in \mathcal{V}
\end{cases}
\end{equation}

With the generalized path-integral formulation, a material-form reparameterization can be performed similarly, with the reference surface union \( \mathcal{B} \) now being a union of reference surfaces \( \mathcal{B_M} \) and interior volumes \( \mathcal{B_V} \). The definition of \( \hat{\Omega} \), \( \D\mu \), and \( J(p_n) \) in \( \hat{f}(\Bar{p}) \) changes accordingly. Then, by applying Reynolds transport theorem, a generalized differential path integral can be obtained: 
\begin{equation}
\label{eq:diff_path_int_volume}
\frac{\partial I}{\partial \pi} = \intbox{\int_{\hat{\Omega}} \frac{\partial \hat{f}(\Bar{p})}{\partial \pi} \D\mu(\Bar{p})} + \bndbox{\int_{\partial \hat{\Omega}} \Delta \hat{f_K}(\Bar{p}) V_K(p_K) \D\mu'(\Bar{p})}.
\end{equation}
This equation has the exact same form as Eq.~\ref{eq:diff_path_int}, but now the two vertices of the boundary segment \( (p_{K-1}, p_K) \) can reside on a surface or in a volume. This makes a difference in the definition of \( \partial\hat{\Omega} \), \( V_K(p_K) \), and \( \D\mu' \). We refer readers to the original paper \cite{Zhang:2021:PSDR} and Zhang's thesis \cite{Zhang:2022:thesis} for more details.

\subsubsection{PSDR of implicit surfaces}

\setlength{\resLen}{1.4in}
\begin{figure}[t]
    \centering
    \small
    \addtolength{\tabcolsep}{-3pt}
    \begin{tabular}{cc}
         \includegraphics[width=\resLen]{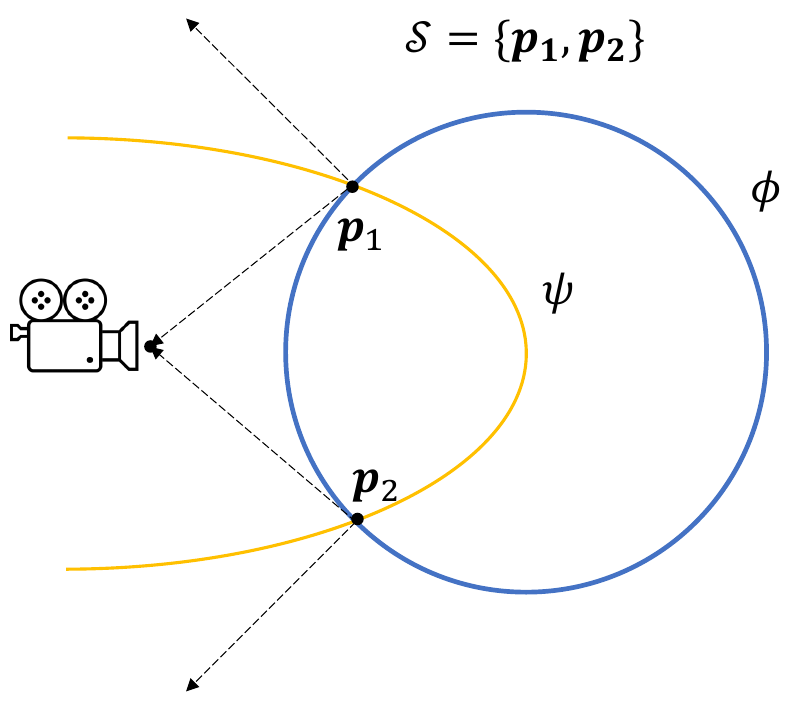} &
         \includegraphics[width=\resLen]{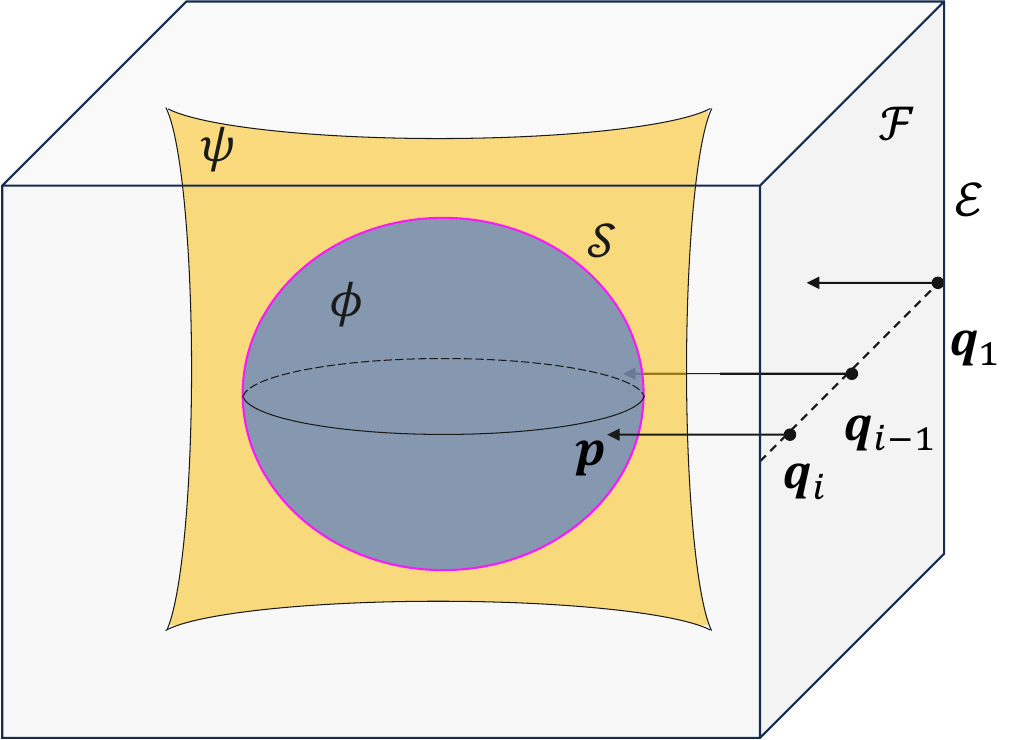} \\
         (a) & (b)
    \end{tabular}
    
    \caption{\label{fig:psdr_sdf}
        \textbf{PSDR for implicit surfaces:} (a) primary boundaries can be viewed as the intersected points (in 3D, curves) of \(\phi\) and \(\psi\). (b) To sample primary boundaries, we cast rays from \( q_i, \dots, q_i \) until the ray intersects with \( \mathcal{S} \) or \( q_i \) reaches the other end. Reference: Figure 5 in \cite{Zhou:2024:PSDR-SDF}.
    }
\end{figure}

Apart from mesh-based representations, implicit surfaces (e.g., SDFs) are also popular in inverse rendering studies, with the advantage in handling topology changes. Therefore, Zhou et al.~\cite{Zhou:2024:PSDR-SDF} have extended PSDR to handle implicit surfaces as well. 
An implicit surface can be defined as zero-level sets of some function \( 
\phi \): \( \{ x \in \mathbb{R}^3, \phi(x,\pi)=0 \} \). The main difference between handling meshes and such implicit surfaces is that these surfaces do not offer a finite set of edges to sample from, and thus require additional steps to sample the visibility boundaries. 

For primary boundaries that occur on the last segment of the light path \( (p_{N-1}, p_N) \) (i.e., the segment that connects to the detector), it can be viewed as an intersected curve \( \mathcal{S} \) of the implicit surface and another implicit surface \( \psi(p) = \langle \nabla_\phi (p), (p_N - p) \rangle = 0\), which describes a set of positions where the surface normal is perpendicular to the viewing direction. This is illustrated in Figure~\ref{fig:psdr_sdf}.
To sample \( \mathcal{S} \), Zhou et al. introduce an axis-aligned bounding box (AABB) of \( \phi \), and sample points on an edge \( \mathcal{E} \) of the AABB. From a sampled point \( q \), a ray is shot in the direction of the face normal, to find if it intersects with a point that belongs to \( \mathcal{S} \). If not, \( q \) is marched on the face along the direction perpendicular to the edge and the rays are shot repeatedly until they find a point in \( \mathcal{S} \) or \( q \) reaches the other end. In practice, the AABB can be sub-divided into a tighter bound using range analysis~\cite{Stol:1997:Self}, reducing the overhead of the marching process. 

For secondary boundaries that occur on other parts of the light path, the sampling process is similar to that of the original PSDR that starts from sampling a point on the surface, generates the boundary segment, and completes the light path from both directions. 
The main difference here is that discontinuities do not occur on edges, but on the whole surface, which means the original integration over edges in Eq.~\ref{eq:psdr_rewrite_xw} is now an integration over a surface:
\begin{equation}
\label{eq:psdr_sdf_int}
\int_{\mathcal{B}} h(p) \D A(p) \defeq \int_{\mathcal{B}} \int_{\mathbb{S}} F'(\omega, p) \D \theta(\omega) \D A(p),
\end{equation}
where \( \mathcal{B} \) is the union of (material-form) surfaces, \( \mathbb{S} \) is the set of tangent directions at point \( p \), \( F' \) is the integrand after reparameterization, and \( \theta \) is the angle measure. However, since sampling points on the surface directly is not feasible for implicit surfaces, Zhou et al. propose to change the integration domain to one of the faces \( \mathcal{F} \) of its AABB.

Given a point on \( \mathcal{F} \), we can cast a ray along the face normal from this point to retrieve the intersection points with the surface. There might be multiple intersection points, and therefore the surface can be divided into multiple domains \( \mathcal{B}_i \) accordingly, where \( i \) indicates the number of intersections encountered by the ray when reaching this domain. This enables the decomposition of the original integral into multiple integral over each domain:
\begin{equation}
\label{eq:psdr_sdf_divide}
\int_{\mathcal{B}} h(p) \D A(p) = \sum_{i=1}^N \int_{\mathcal{B}_i} h(p) \D A(p),
\end{equation}
where \( N \) is the maximum possible number of intersections a ray can encounter.
Besides, this ray casting process naturally defines a mapping \( \mathcal{T}_i \) from a point \( q \) on \( \mathcal{F} \) to a point \( p \) on the surface domain \( \mathcal{B}_i \), which means we can apply a change of variable to the integration. Note that the mapping \( \mathcal{T}_i \) might not involve all the points on \( \mathcal{F} \), since not all rays have the \(i\)-th intersection point. In this case, we change the original integrand into a new one:
\begin{equation}
\label{eq:psdr_sdf_integrand}
g_i(q) \defeq \\
\begin{cases}
h(\mathcal{T}_i(q)) J_{\mathcal{T}_i}(q), & \mathcal{T}_i(q)  \; \text{exists} \\
0, & \text{otherwise}
\end{cases}
\end{equation}
where \( J_{\mathcal{T}_i} \) is the Jacobian determinants of \( \mathcal{T}_i \), and the integration can be expressed as:
\begin{equation}
\label{eq:psdr_sdf_final}
\int_{\mathcal{B}} h(p) \D A(p) = \sum_{i=1}^N \int_{\mathcal{F}} g_i(q) \D A(q).
\end{equation}

\subsection{Reparameterization of boundary integral}
\label{sec:theory_was}

Apart from the strategies that directly identify the discontinuity boundaries and estimate their contributions, there exists another route that tries to use reparameterization to avoid explicit sampling of the discontinuities in the integrand and transform the boundary integral into an interior integral. This essentially means that we no longer need to decompose the differentiation into two terms and sample the edges explicitly. 

\subsubsection{Reparameterization with spherical rotations}

The first approach of this idea is introduced by Loubet et al.~\cite{Loubet:2019:Reparameterizing}, which reparameterizes the spherical integral in Eq.~\ref{eq:diff_RE_int} using spherical rotations to avoid explicit sampling of the discontinuities. A spherical rotation \( R \) is a transformation that rotates a unit direction \( \omega \) to another unit direction \( \omega' = R (\omega, \pi) \) by a fixed angle. This transformation enables a reparameterization of the integral in Eq.~\eqref{eq:diff_RE_int}:
\begin{equation}
\label{eq:was_reparam_int}
I = \int_{\mathbb{S}^2} f(\omegai) \D \sigma(\omegai) = \int_{\mathbb{S}^2} f(R(\omega, \pi)) \D \sigma(\omega),
\end{equation}
where \( f \defeq L_if_s \) is the original integrand. Now suppose \( f(\omega) \) has a discontinuous point at \( \omega = T(\pi) \) that is dependent on \( \pi \). For \( g(\omega) \defeq f(R(\omega, \pi)) \), there also exists a discontinuous point \( \omega = \omega_0 \) satisfying \( R(\omega_0, \pi) = T(\pi) \). When \( R \) is carefully chosen to approximate the motion of the discontinuity, 
that is, \( \partial R / \partial \pi = \partial T / \partial \pi \), 
the discontinuity point \(\omega_0\) becomes \( \pi \)-independent. 
In this case, it holds that
\begin{equation}
\label{eq:was_reparam_diff}
\frac{\partial I}{\partial \pi}
= \frac{\partial}{\partial \pi} \int_{\mathbb{S}^2} g(\omega) \D \sigma(\omega) = \intbox{ \int_{\mathbb{S}^2} \frac{\partial}{\partial \pi} g(\omega) \D \sigma(\omega) },
\end{equation}
which means we no longer need to sample and estimate the boundary integral.

To construct the correct \( R \), Loubet et al. propose to cast a few auxiliary rays that are centered around the primal ray during each light bounce, to approximate the motion of the discontinuity boundary. If there exists an evolving discontinuity boundary around the primal ray, the auxiliary rays should catch the motion of the object to which the boundary belongs. If there are multiple rays that catch the motion, one of them is selected using a heuristic that best approximates the boundary's motion. This motion is then used to construct \( R \) such that the motion \( R \) given to the direction of the auxiliary ray matches the motion of the boundary. 

\subsubsection{Warped-area reparameterization}

\setlength{\resLen}{1.6in}
\begin{figure}[t]
    \centering
    \small
    \addtolength{\tabcolsep}{-3pt}
    \begin{tabular}{cc}
         \includegraphics[trim={0 0 0 0},clip, width=\resLen]{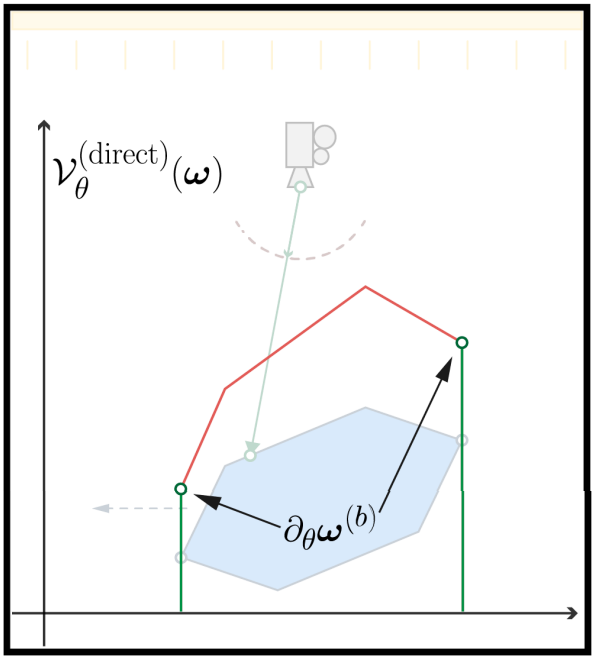} &
         \includegraphics[trim={0 0 0 0},clip, width=\resLen]{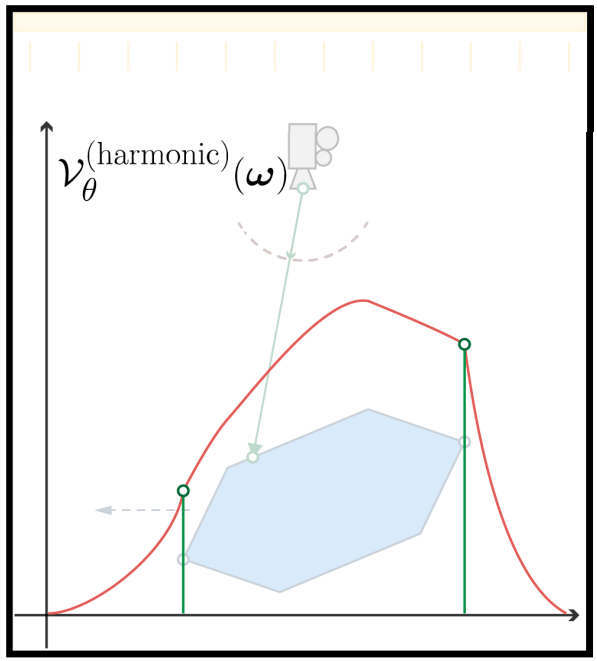} \\
         (a) & (b)
    \end{tabular}
    
    \caption{\label{fig:was}
        \textbf{Constructing warp field:} (a) a discontinuous warp field constructed by tracing auxiliary rays. (b) A continuous warp field smoothed by the convolution kernel \( w \). Note that the values on boundary remain intact. Reference: Figure 6 in \cite{bangaru:2020:warpedsampling}.
    }
\end{figure}

Although the above approach successfully avoids sampling the discontinuities explicitly, it is biased due to the approximation of boundary's motion using one of the auxiliary rays. To this end, Bangaru et al.~\cite{bangaru:2020:warpedsampling} propose warped-area sampling (WAS), a refined method that offers unbiased estimation of the differentiation. 

Warped-area sampling draws inspiration from the divergence theorem:
\begin{equation}
\label{eq:was_divergence}
\int_{\partial\Omega} F \cdot n^\partial \D \ell = \int_\Omega \nabla \cdot F \D \sigma,
\end{equation}
where \( F \) is a continuous vector field defined on \( \Omega \), \( n \) is the normal of the boundary \( \partial \Omega \), and \( \nabla \cdot F \) is the divergence of \( F \). This theorem essentially reparameterizes a boundary integral to an interior integral that no longer requires explicit boundary sampling, and when applied to the boundary integral in Eq.~\ref{eq:diff_RE_Reynolds}, we have
\begin{equation}
\label{eq:was_divergence_boundary}
\bndbox{\int_{\Delta\mathbb{S}^2} (\Delta f v^\partial) \cdot n^\partial \D \ell(\omegai) } = \intbox{\int_{\mathbb{S}^2} (\nabla \cdot \Delta f \mathcal{V}) \D \sigma(\omegai)},
\end{equation}
where \( \mathcal{V} \) is called a \textit{warp field} defined on \( \mathbb{S}^2 \) instead of its discontinuity boundaries only. By definition, this warp field \( \mathcal{V} \) must satisfy: (a) \( \mathcal{V} \) is continuous, and (b) when \( \omega \) approaches the boundary \( \Delta \mathbb{S}^2 \), \( \mathcal{V}(\omega) \) still converges to \( v^\partial \). Therefore, the problem comes down to the construction of such a warp field. Note that the reparameterization using spherical rotation \( R \) proposed by Loubet et al. \cite{Loubet:2019:Reparameterizing} can actually be interpreted as constructing an \textit{imperfect} warp field \( \mathcal{V} = \partial R / \partial \pi \) that approximates (but does not strictly satisfy) requirement (b), or the boundary requirement. On the other hand, a warp field can also be converted to a transformation. This is more thoroughly discussed in Appendix C of the original paper \cite{bangaru:2020:warpedsampling}. 

The construction of a warp field that satisfies both requirements can be broken down into two steps: first, construct a warp field \( \mathcal{V}^{\text{dis}} \) whose boundary is the same as \( v^\partial \) but discontinuous, and second, smooth the warp field into a continuous field. This is illustrated in Figure~\ref{fig:was}. The first step is achieved by casting auxiliary rays around the primal sampled ray. If a ray with direction \( \omegap \) intersects with an object dependent on \( \pi \), we can obtain \( \vp \defeq {\partial \omegap} / {\partial \pi} \) by differentiating the ray intersection process. This gives the definition of velocity in the interior of the warp field, while ensuring that as \( \omegap \) approaches the boundary \( \Delta \mathbb{S}^2 \), its velocity \( \vp \) converges to \( v^\partial \). 

The warp field in the first step is discontinuous at its boundary, since the rays cast on different sides of the boundary usually intersect with different objects, resulting in different velocities. Therefore, we need a smoothing step to create a continuous field, while preserving the velocity \( v^\partial \) at the boundary. This is achieved using a convolution over the warp field:
\begin{equation}
\label{eq:was_conv}
\mathcal{V}(\omega) \defeq \frac{\int_{\mathbb{S}^2} w(\omega, \omega')\mathcal{V}^{\text{dis}} (\omega') \D \sigma(\omega')}{\int_{\mathbb{S}^2} w(\omega, \omega') \D \sigma(\omega')}.
\end{equation}
where the convolution kernel is expressed as
\begin{equation}
\label{eq:was_conv_kernel}
w(\omega, \omega') \defeq \frac{1}{\mathcal{D}(\omega, \omega') + \mathcal{B}(\omega')},
\end{equation}
where \( \mathcal{D}(\omega, \omega') \) is a distance function, and \( \mathcal{B}(\omega') \) is a boundary-test function that converges to \( 0 \) when \( \omega' \rightarrow \Delta\mathbb{S}^2 \). 

After constructing the warp field \( \mathcal{V} \), we can apply Eq.~\ref{eq:was_divergence_boundary} and reparameterize the boundary integral into an interior integral. Therefore, the two integrals in Eq.~\ref{eq:diff_RE_Reynolds} can now be merged into one interior integral, known as the \textit{reparameterized spherical integral}:
\begin{equation}
\label{eq:was_reparam_spherical_int}
\frac{\partial I}{\partial \pi} = \intbox{ \int_{\mathbb{S}^2} \left[ \frac{\partial f}{\partial \pi} + (\nabla \cdot \Delta f \mathcal{V}) \right] \D\sigma(\omegai) },
\end{equation}
which can be estimated without the need of explicit edge sampling.

\subsubsection{Warped-area reparameterization for differential path integral}

\setlength{\resLen}{3.5in}
\begin{figure}[t]
    \centering
    \small
    \addtolength{\tabcolsep}{-3pt}
    \begin{tabular}{c}
         \includegraphics[trim={25 12 0 0},clip, width=\resLen]{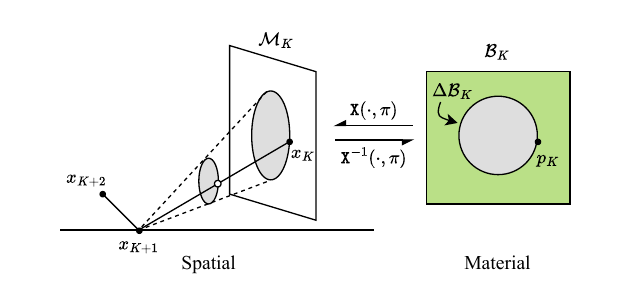}
    \end{tabular}
    
    \caption{\label{fig:was_psdr}
        \textbf{Warped-area reparameterization for differential path integral:} this scene illustrates the boundary segment \( (x_K, x_{K+1}) \) and the corresponding discontinuity curve \( \Delta \mathcal{B}_K \). After applying divergence theorem, the domain of integration is converted to the whole plane \( \mathcal{B}_K \). Reference: Figure 6 in \cite{Xu:2023:PSDR-WAS}.
    }
\end{figure}

Similar to edge sampling, WAS is performed on the spherical integral involving a single light bounce. This means the warp field needs to be constructed repeatedly for every light bounce, which is only applicable to basic rendering algorithms such as unidirectional path tracing. To address this, Xu et al.~\cite{Xu:2023:PSDR-WAS} introduce the idea of warped-area reparameterization to differential path integral. 

Starting from Eq.~\ref{eq:diff_path_int} in PSDR, the boundary integral can be decomposed into a sum of integrals:
\begin{equation}
\label{eq:was_psdr_boundary}
\begin{split}
& \int_{\partial \hat{\Omega}} \Delta \hat{f_K}(\Bar{p}) V_K(p_K) \D\mu'(\Bar{p}) \\ 
&= \sum_{N=1}^{\infty} \sum_{K=0}^{N-1} \int_{\partial \hat{\Omega}_{N,K}} \Delta\hat{f}_K(\Bar{p}) V_K(p_K) \D\mu'(\Bar{p}) \\
&= \sum_{N=1}^{\infty} \sum_{K=0}^{N-1} \int_{\mathcal{B}^N} \bigg[ \underbrace{ \int_{\Delta \mathcal{B}_K} \Delta\hat{f}_K(\Bar{p}) V_K(p_K) \D\ell(p_K) 
}_{\eqdef I_{N,K}} \bigg] \prod_{n \neq K} \D A(p_n),
\end{split}
\end{equation}
where \( \partial \hat{\Omega}_{N,K} \) is a sub-space of \( \partial \hat{\Omega} \) which only includes light paths with \( N \) segments and \( (p_K, p_{K+1}) \) is a boundary segment, and \( \Delta \mathcal{B}_K \) is the set of visibility boundaries of the (material form) surface \( \mathcal{B}_K \) that \( p_K \) resides on. Figure~\ref{fig:was_psdr} illustrates a simple scene containing a boundary segment. Note that the boundary segment is defined at \( (p_K, p_{K+1}) \), which is slightly different from \( (p_{K-1}, p_K) \) in the original PSDR. This is for the simplicity of annotation in the following equations. After this decomposition, it is now possible to reparameterize the inner integral \( I_{N,K} \) using divergence theorem:
\begin{equation}
\label{eq:was_psdr_divergence}
\begin{split}
& I_{N,K} = \\
& \bndbox{ \int_{\Delta \mathcal{B}_K} \big( \Delta\hat{f}_K v_K^\partial \big) \cdot n_K^\partial \D\ell(p_K) } 
= \intbox{ \int_{\mathcal{B}_K} \big( \nabla \cdot \Delta\hat{f}_K \mathcal{V}_K \big) \D A(p_K) }.
\end{split}
\end{equation}
Merging Eq.~\ref{eq:was_psdr_divergence} back to Eq.~\ref{eq:was_psdr_boundary} yields the reparameterized boundary integral:
\begin{equation}
\label{eq:was_psdr_boundary_reparam}
\begin{split}
& \int_{\partial \hat{\Omega}} \Delta \hat{f_K}(\Bar{p}) V_K(p_K) \D\mu'(\Bar{p}) \\ 
&= \sum_{N=1}^{\infty} \sum_{K=0}^{N-1} \int_{\mathcal{B}^{N+1}} \left[ \nabla \cdot \Delta\hat{f}_K \mathcal{V}_K \right] (p_K) \D\mu(\Bar{p}) \\
&= \int_{\hat{\Omega}} \left[ \sum_{K=0}^{N-1} \left[ \nabla \cdot \Delta\hat{f}_K \mathcal{V}_K \right] (p_K) \right] \D\mu(\Bar{p}),
\end{split}
\end{equation}
which is now performed on the path space \( \hat{\Omega} \) instead of the boundary path space \( \partial \hat{\Omega} \). This means the interior integral and boundary integral in Eq.~\ref{eq:diff_path_int} can be composed into one interior integral, known as the \textit{reparameterized differential path integral}:
\begin{equation}
\label{eq:was_psdr_reparam_diff_path_int}
\frac{\partial I}{\partial \pi} = \intbox{ \int_{\hat{\Omega}} \left[ \frac{\partial \hat{f}(\Bar{p})}{\partial \pi} + \sum_{K=0}^{N-1} \left[ \nabla \cdot \Delta\hat{f}_K \mathcal{V}_K \right] (p_K) \right] \D\mu(\Bar{p}) }.
\end{equation}

Compared to the reparameterized spherical integral in Eq.~\ref{eq:was_reparam_spherical_int}, the path integral in Eq.~\ref{eq:was_psdr_reparam_diff_path_int} has a more generalized form that depicts multi-bounce light transport within a scene. This enables more advanced Monte Carlo estimation strategies targeting path sampling, e.g., bidirectional path tracing, which is essential to achieve low variance in complex scenes. 

\subsubsection{Reparameterization for differentiable SDF rendering}

There exist two concurrent works that utilize reparameterization in differentiable SDF rendering. Vicini et al.~\cite{Vicini:2022:DiffSDF} build upon the idea of \cite{Loubet:2019:Reparameterizing} that uses a reparameterization function, while Bangaru et al.~\cite{Bangaru:2022:DiffNeuralSDF} build upon the idea of \cite{bangaru:2020:warpedsampling} that constructs a warp field. However, as mentioned earlier, these two methods are interrelated and interpretable to each other. Another difference is that \cite{Vicini:2022:DiffSDF} uses a voxel grid as the underlying representation of the SDF, while \cite{Bangaru:2022:DiffNeuralSDF} uses a neural network. In this paper, we will introduce the reparameterization method proposed by Vicini et al. \cite{Vicini:2022:DiffSDF}.

A signed-distance function \( \phi(x) \) measures the distance of a point \( x \) to its surface, and the distance is negative when the point is inside the surface. Since SDFs have continuous surfaces and provide analytical equations for them, it is actually easier to construct a reparameterization function \( \mathcal{T} \) that strictly satisfies the boundary requirement proposed in \cite{bangaru:2020:warpedsampling}. The boundary requirement can be expressed as 
\begin{equation}
\label{eq:was_sdf_boundary_req}
\partial_\pi \mathcal{T}(\omegab, \pi) = \partial_\pi \omegab,
\end{equation}
for all \( \omegab \) lying on a discontinuity boundary on \( \mathbb{S}^2 \). It states that the motion after reparameterization should strictly match the motion of the original boundary. Note that for differential SDFs, we use the spherical integral form of the rendering equation, as shown in Eq.~\ref{eq:was_reparam_int}. In order to construct a reparameterization \( \mathcal{T} \), we can first construct a 3D vector field \( \mathcal{V} \):
\begin{equation}
\label{eq:was_sdf_warp_field}
\mathcal{V}(x,\pi) \defeq - \frac{\partial_x \phi(x,\pi_0)}{\Vert \partial_x \phi(x,\pi_0) \Vert^2} \phi (x,\pi),
\end{equation}
where \( x \in \mathbb{R}^3 \) is any point in the 3D space, and \( \pi_0 \) is the value of \( \pi \) in the current scene, which could also be seen as a "detached" version of \( \pi \). It can be proved that for points on the surface, \( \partial_\pi \mathcal{V}(x, \pi) = \partial_\pi x(\pi) \), where \( x \) is an intersected point on the surface through ray tracing, which makes it dependent on \( \pi \).
Now, we can define the reparameterization
\begin{equation}
\label{eq:was_sdf_reparam}
\mathcal{T}(\omega, \pi) \defeq \frac{\bar{\mathcal{T}}(\omega, \pi)}{\Vert \bar{\mathcal{T}}(\omega, \pi) \Vert},
\end{equation}
with
\begin{equation}
\label{eq:was_sdf_aux}
\bar{\mathcal{T}}(\omega, \pi) \defeq t\omega + \mathcal{V}(x_t, \pi) - \mathcal{V}(x_t, \pi_0),
\end{equation}
where \( x_t \defeq x_0 + t\omega \) is the target position of a ray cast from \( x_0 \) with direction \( \omega \) and travel distance \( t \). Here, \( t \) is an evaluation distance function, which converges to the intersection distance when the intersected point on the SDF is approaching a visibility boundary, and therefore \( x_t \) is not necessarily on the surface of the SDF. Subtracting \( \mathcal{V}(x_t, \pi_0) \) is to ensure that the reparameterization is an identity map when \( \pi = \pi_0 \), but the derivative is changed. Now, the derivative becomes
\begin{equation}
\label{eq:was_sdf_reparam_diff}
\partial_\pi \mathcal{T}(\omega, \pi) = \frac{1}{t} (\mathbb{I} - \omega \cdot \omega^T) \partial_\pi \mathcal{V}(x_t, \pi),
\end{equation}
where \( \mathbb{I} - \omega \cdot \omega^T \) is projecting the gradient of the vector field to the unit sphere's tangent space, and \( 1 / t \) is scaling the motion from distance \( t \). This reparameterization ensures that the motion of the discontinuities is correctly captured. 

The choice of the evaluation distance function \( t \) is also important to the reparameterization. It can be constructed by reusing the samples when performing sphere tracing \cite{Hart:1996:Sphere} to compute the intersection of the ray with SDF:
\begin{equation}
\label{eq:was_sdf_t}
t \defeq \frac{\sum_{i=1}^N w(i)t_i}{\sum_{i=1}^N w(i)},
\end{equation}
where \( t_i \) are the intermediate distances obtained when performing sphere tracing, and \( w \) is a weighting function:
\begin{equation}
\label{eq:was_sdf_w}
w(i) \defeq w_\mathrm{edge}(i) w_\mathrm{dist}(i) w_\mathrm{bbox}(i).
\end{equation}
The general idea of this weight is to ensure that \( t \) converges to the intersection distance when approaching a visibility boundary, while keeping the variance low. We refer readers to Section 4.2 of the original paper \cite{Vicini:2022:DiffSDF} for more detailed discussion of each term.

\section{Monte Carlo Estimation of the Interior Term}
\label{sec:interior}

In the previous section, we have introduced three different theories for physics-based differentiable rendering. One thing in common is that they all contain a similar interior term, as shown in the differential rendering equations Eq.~\ref{eq:diff_RE_Reynolds}, Eq.~\ref{eq:diff_path_int}, and Eq.~\ref{eq:was_psdr_reparam_diff_path_int}. This interior term requires an integration of \( \partial f / \partial \pi \) over a certain domain, where \( f \) is the measurement contribution function that often contains a series of multiplications of BSDFs, detector importance, and geometric terms. In most cases, the estimation of this interior integral can be directly handled by auto-diff through a forward rendering and backpropagation pass, which essentially means the samples from forward rendering are "reused" to estimate the interior term. This is not often the best approach, and in the worst case where some part of the function equals zero and is never sampled, it causes severe bias since the derivative of the function might not be zero and needs to be sampled. This is demonstrated and discussed in \cite{Nimier:2022:Volume} in the context of differential volume rendering. 

In practice, since the samples are often drawn by a mapping from uniform samples, this mapping might also be differentiated, when the construction of the mapping depends on \( \pi \). Note that either attaching or detaching this mapping to auto-diff does not affect the unbiasedness of the estimation, but does have an impact on its variance. Zeltner et al.~\cite{Zeltner:2021:Light} provide a thorough study on the impact of such attached or detached sampling strategies on the variance of the estimation. Their study also points out that the samples reused from forward rendering are often not suitable for estimating the derivative, and will lead to high variance in the estimation result. The better approach would be sampling according the derivative with detached mapping. In the following sections, we will present a few works focusing on sampling strategies that are particularly optimized for estimation of the interior term.

\subsection{BSDF}

In forward rendering, BSDF importance sampling is commonly used to reduce variance when the BSDF is glossy or near-specular, which is done by drawing \( \omegai \) the probability density proportional to the BSDF \( \bsdf \). However, this is not suitable for differentiable rendering, as we are now evaluating \( \partial \bsdf / \partial \pi \), which is often largely different from \( \bsdf \). 

\subsubsection{Antithetic sampling}

Antithetic sampling is a commonly used way to reduce variance when the integrand approximates an odd function. When sampling two antithetic samples \( x_1,x_2 \) with \( x_1 + x_2 = 0 \), the resulting estimation also approaches \( 0 \), which offers significantly lower variance. In the field of differentiable rendering, the idea of antithetic BSDF sampling is mentioned in \cite{Loubet:2019:Reparameterizing, Zeltner:2021:Light} and formally discussed by Zhang et al.~\cite{Zhang:2021:Antithetic}, which proves to be a simple yet effective way to reduce variance for estimating the interior integral. 

When applying antithetic sampling to the differentiation of BSDFs, one key observation is that many commonly used glossy BSDFs rely on a normal distribution function (NDF) that is often symmetric. For instance, a microfacet BSDF can be expressed as
\begin{equation}
\label{eq:interior_antithetic_microfacet}
\bsdf(\omegai, \omegao) = D(\omegah)\bsdf^{(0)}(\omegai, \omegao),
\end{equation}
and its derivative
\begin{equation}
\label{eq:interior_antithetic_microfacet_diff}
\frac{\partial \bsdf}{\partial \pi} = \frac{\partial D}{\partial \omegah} 
\cdot \frac{\partial \omegah}{\partial \pi}\bsdf^{(0)} + D\frac{\partial \bsdf^{(0)}}{\partial \pi},
\end{equation}
where \( D \) is the NDF parameterized by the half-way vector \( \omegah \defeq (\omegai + \omegao) / \Vert \omegai + \omegao \Vert \), and \( \bsdf^{(0)} \) is the other terms such as Fresnel reflection. When computing the derivative, the first term often dominants the result as \( D \) varies more rapidly, especially if the BSDF is glossy. When the NDF is symmetric, which is often the case such as Beckmann or GGX models, \( \partial D / \partial \omegah \) becomes an odd function, which enables the antithetic sampling strategy. When drawing the sample from the BSDF, a half-way vector \( \omega_\mathrm{h,1} \) is first drawn based on the NDF, followed by another half-way vector \( \omega_\mathrm{h,2} \) that is set to be symmetric to \( \omega_\mathrm{h,1} \) w.r.t. z-axis (under local coordinate). Then, the other terms including \( \omega_\mathrm{i,1} \) and \( \omega_\mathrm{i,2} \) are computed accordingly. 

For a single light bounce, this strategy produces two samples that eventually branch the light path into two paths that both contribute to the estimation. However, when there are multiple glossy surfaces in the scene, applying antithetic sampling to every light bounce on such surfaces is not feasible, as the number of light paths will grow exponentially. Therefore, another way would be only applying antithetic sampling once for a given light path, and merging the branched paths back after a few bounces (when the intersected surface is no longer glossy). This light path is then repeated \( k \) times to apply antithetic sampling to \( k \) different glossy surfaces that the light path encounters. 

\subsubsection{Decomposed sampling}

Antithetic sampling is a simple approach to reduce variance in many cases, but it is not guaranteed to be effective in all cases. Belhe et al.~\cite{Belhe:2024:BRDF} point out that there exist BSDF derivatives that are even functions, e.g., w.r.t. the roughness of GGX, in which case antithetic sampling produces high variance. Therefore, their study dives deeper into specialized strategies for sampling differential BSDFs. 

Given a BSDF \( \fs \) and a scene parameter \( \pi \), the ideal sampling strategy that could produce zero variance would be sampling a distribution proportional to the integrand \( \partial \fs / \partial \pi \). This is also proposed as differential BSDF sampling in \cite{Zeltner:2021:Light}. However, this sampling strategy only produces zero variance under the condition that \( \partial \fs / \partial \pi \) is non-negative, which is not always the case. When the integrand has negative parts, it becomes impossible to produce zero variance through a single sampling distribution, which is also known as sign variance. As a side note, in practice the integrand does not contain only a single BSDF term, and therefore the zero variance discussion is only with regard to this single term, while other terms will still produce variance.

To address the problem of sign variance, the first strategy introduced in \cite{Belhe:2024:BRDF} is called positivization \cite{Owen:2000:Safe}, which deals with sign variance by analytically dividing \( \partial \fs / \partial \pi \) into positive and negative terms, and sampling them separately. This is very effective and is demonstrated in the original paper \cite{Belhe:2024:BRDF} when applied to isotropic GGX models. However, it is not guaranteed that we can analytically obtain the roots of \( \partial \fs / \partial \pi \) for every BSDF, thus limiting the application of this strategy. 

The second strategy is called product decomposition, building upon the observation that some BSDFs can be expressed as the product of two functions:
\begin{equation}
\label{eq:interior_bsdf_product}
f(\omegah, \pi) = N(\pi) g(\omegah, \pi),
\end{equation}
where \( g(\omegah, \pi) \) is a non-negative shape function, and \( N(\pi) \) is a normalization term independent of \( \omegah \). And for its derivative,
\begin{equation}
\label{eq:interior_bsdf_product_diff}
\partial_\pi \f(\omegah, \pi) = \partial_\pi N(\pi) g(\omegah, \pi) + N(\pi) \partial_\pi g(\omegah, \pi),
\end{equation}
the first term is single-signed for \( \omegah \), while the second term might not be. Fortunately, for a set of BSDFs, including anisotropic microfacet BRDFs, diffuse BSSRDFs, and isotropic ABC BRDFs, the second term is also single-signed, which enables these BSDFs to be sampled with the product decomposition strategy and achieve low variance. 

The last strategy is called mixture decomposition, targeting the situation when the parameter \( \pi \) is acting as a combination weight in the BSDF:
\begin{equation}
\label{eq:interior_bsdf_mixture}
f(\omegai, \omegao, \pi) = w_1(\pi)f_1(\omegai, \omegao) + w_2(\pi)f_2(\omegai, \omegao),
\end{equation}
and
\begin{equation}
\label{eq:interior_bsdf_mixture_diff}
\partial_\pi f(\omegai, \omegao, \pi) = \frac{\partial w_1}{\partial \pi}f_1(\omegai, \omegao) + \frac{\partial w_2}{\partial \pi}f_2(\omegai, \omegao),
\end{equation}
where \( f_1 \) and \( f_2 \) are different lobes that are non-negative, and \( w_1 \) and \( w_2 \) are their respective weights. In this case, since \( w_1 \) and \( w_2 \) do not change with \( \omegai \) and \( \omegao \), the two terms of \( \partial_\pi f(\omegai, \omegao, \pi) \) are again single-signed and can be sampled separately. This strategy covers BSDFs that have such a mixture structure, such as all Uber BRDFs.

\subsection{Pixel reconstruction filter}

\setlength{\resLen}{1.6in}
\begin{figure}[t]
    \centering
    \small
    \addtolength{\tabcolsep}{-3pt}
    \begin{tabular}{cc}
         \includegraphics[trim={40 5 40 10},clip, width=\resLen]{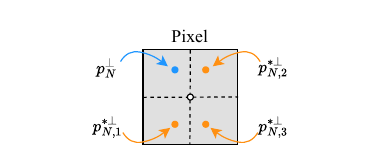} &
         \includegraphics[trim={40 15 10 0},clip, width=\resLen]{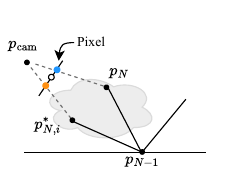} \\
         (a) & (b)
    \end{tabular}
    
    \caption{\label{fig:interior_pixel}
        \textbf{Pixel reconstruction filter:} (a) we use antithetic sampling to obtain 3 additional samples \( p_{N,i}^{*\perp} \) for one pixel on the image plane. (b) The samples are used to cast rays to the scene to retrieve \( p_{N,i}^* \). Reference: Figure 3 in \cite{Yu:2022:Pixel}.
    }
\end{figure}

In addition to antithetic BSDF sampling \cite{Zhang:2021:Antithetic}, Yu et al.~\cite{Yu:2022:Pixel} also propose to leverage antithetic sampling on differentiating the pixel reconstruction filter of a pinhole camera. Following the method in \cite{Zhang:2021:PSDR}, pinhole cameras can be supported in PSDR by encoding the detector importance:
\begin{equation}
\label{eq:interior_antithetic_pinhole}
\We (p_{N-1} \rightarrow p_N) \defeq \fv(p_{N-1} \rightarrow p_{N} \rightarrow \pcam) \frac{G(p_N \leftrightarrow \pcam) h(p_N^\perp)}{(\ncam \cdot \overrightarrow{\pcam p_N})^3},
\end{equation}
where \( \pcam \) is the (material form) position or center of projection of the pinhole camera, \( \ncam \) is its axis of projection, \( h \) is the pixel reconstruction filter, \( p_N^\perp \) is the projection of \( p_N \) on the image plane, and \( \overrightarrow{\pcam p_N} \) is the unit vector from \( \pcam \) to \( p_N \). Through this encoding, we can avoid introducing additional Dirac delta functions in the derivative due to this zero-measure detector. 

When there exists discontinuity boundary that is directly visible to the camera, \( h(p_N^\perp) \) becomes dependent on \( \pi \). And when estimating \( \partial h / \partial \pi \), one key observation is that \( h(p_N^\perp) \) often exhibits point symmetry with respect to the pixel center. Therefore, antithetic sampling can be utilized to reduce the variance of this estimation. Given a sampled light path \( \bar{p} = (p_0, \cdots, p_{N-1}, p_N) \) with probability density function \( \pdf(\bar{p}) \), we can construct 3 additional light paths \( \bar{p}_i^* = (p_0, \cdots, p_{N-1}, p_{N,i}^*) \) with \( i = 1,2,3 \), such that the four points on the image plane \( p_N^\perp, p_{N,1}^{*\perp}, p_{N,2}^{*\perp}, p_{N,3}^{*\perp}  \) are symmetric w.r.t. the pixel center. 

In practice, we first obtain \( p_N^\perp \) and then \( p_{N,i}^{*\perp} \) on the image plane. Then, rays are cast from the center through \( p_{N,i}^{*\perp} \) and intersect with the scene to obtain \( p_{N,i}^* \). Figure~\ref{fig:interior_pixel} illustrates this sampling process. 
The probability density function \( \pdf_i^*(\bar{p}_i^*) \) for the three antithetic samples can also be computed analytically based on \( \pdf(\bar{p}) \). Finally, the four samples are combined using multiple importance sampling (MIS) \cite{veach1997robust}. Note that we are currently under the assumption that the pixel filter is continuous, and therefore we are only dealing with the interior term. For pixel reconstruction filters that are not continuous (e.g., box filtering), the derivative of the rendering equation will contain an additional boundary term that needs to be addressed. We refer readers to Section 6 of the original paper \cite{Yu:2022:Pixel} for more information.

\subsection{Efficient Differentiation}
In the previous sections, we relied on automatic differentiation (auto-diff) for gradient computation. However, directly applying auto-diff to the aforementioned Monte Carlo estimators can introduce substantial computational overhead and memory consumption. Radiative Backpropogation (RB) \cite{Nimier:2020:Radiative} and Path Replay Backpropogation (PRB) \cite{Vicini:2021:PathReplay} were introduced to address this issue. We will focus on PRB due to its superior efficiency and unbiasedness. In addition, we will follow Zhang~\cite[chapter~7]{Zhang:2022:thesis} and reformulate it using PSDR as it offers simpler derivation. The notations of surface-only light transport from \S\ref{sec:theory_psdr} will be used.  

We want to use auto-diff to compute the interior term in material form from Eq.~\ref{eq:diff_path_int}:
\begin{equation}
    \frac{\partial I}{\partial \pi} = \int_{\hat{\Omega}} \frac{\partial \hat{f}(\Bar{p})}{\partial \pi} \D\mu(\Bar{p}).
\end{equation}
Notice that the Jacobian $\frac{\partial I}{\partial \pi}$ is an enormous matrix, with dimensions determined by the number of pixels and the number of scene parameters being differentiated. Storing this matrix alone requires substantial memory, and computing it via reverse-mode auto-diff demands even more storage for a larger computational graph, capturing all the paths used to estimate the integral. Fortunately, inverse rendering problems are typically formulated as follows:
\begin{equation}
    \pi^*=\arg \min _{\pi} \mathcal{L}(I(\pi)).
\end{equation}
Thus, we aim to compute $\frac{\partial \mathcal{L}}{\partial \pi}$ instead, which is the gradient of a scalar-valued loss function $\mathcal{L}(I)$ with respect to scene parameters $\pi$. Leveraging the material form parameterization, we can derive another interior path integral using the chain rule:
\begin{equation}
\label{eq:diff_image_loss_int}
    \frac{\partial \mathcal{L}}{\partial \pi}
    = \frac{\partial \mathcal{L}}{\partial I} \frac{\partial I}{\partial \pi} = \int_{\hat{\Omega}} \frac{\partial \mathcal{L}}{\partial I} \frac{\partial \hat{f}(\Bar{p})}{\partial \pi} \D\mu(\Bar{p}),
\end{equation} and avoids explicitly computing or storing $\frac{\partial I}{\partial \pi}$. 

However, evaluating $\frac{\partial \hat{f}(\Bar{p})}{\partial \pi}$ can still lead to a large computational graph when the number of vertices in $\Bar{p}$ is high. Vicini et al.~\cite{Vicini:2021:PathReplay} address this issue by utilizing pseudorandom number generators to "replay" the path on the fly, reusing the random seed from the primal rendering. This approach eliminates the need for path recording while ensuring unbiased gradient computation. We illustrate this concept again using the material path formulation. Let
\begin{equation}
g_n(\Bar{p}) = 
\begin{cases}
    \fv(\Bar{x}, n)J(p_n), & 0 \leq n \leq N \\
    G(x_{n-N-1} \leftrightarrow x_{n - N}), & N < n \leq 2N \\
\end{cases}
\end{equation}
Then we can use product rule to transform $\frac{\partial \hat{f}(\Bar{p})}{\partial \pi}$ from Eq.~\ref{eq:diff_image_loss_int} into
\begin{equation}
\begin{split}
    \frac{\partial \hat{f}(\Bar{p})}{\partial \pi} 
    &= \frac{\partial}{\partial \pi} \left( \prod_n g_n(\Bar{p}) \right) \\
    &= \sum_n \frac{\partial g_n(\Bar{p})}{\partial \pi} \prod_{n'\neq n} g_{n'}(\Bar{p}) \\
    &= \sum_n \frac{\partial g_n(\Bar{p})}{\partial \pi} \frac{\hat{f}(\Bar{p})}{g_n(\Bar{p})}.
\end{split}
\end{equation}
Through this transformation, we notice that $\frac{\partial g_n(\Bar{p})}{\partial \pi}$ is the only component that requires auto-diff, which greatly reduces the size of the computational graph. This form also requires less storage, as
$g_n(\Bar{p})$ can be easily obtained by ``replaying'' $\Bar{p}$ using the same random seed and we only need to store $\hat{f}(\Bar{p})$.

For the original formulation of PRB, we refer readers to \cite{Vicini:2021:PathReplay}, where PRB is also generalized to specular materials and volume transport. \cite{Zhang:2022:thesis} discusses additional strategies to efficiently differentiate integrals in PSDR.

\section{Monte Carlo Estimation of the Boundary Term}
\label{sec:boundary}

\setlength{\resLen}{3.2in}
\begin{figure}[t]
    \centering
    \small
    \addtolength{\tabcolsep}{-3pt}
    \begin{tabular}{c}
         \includegraphics[trim={0 0 0 0},clip, width=\resLen]{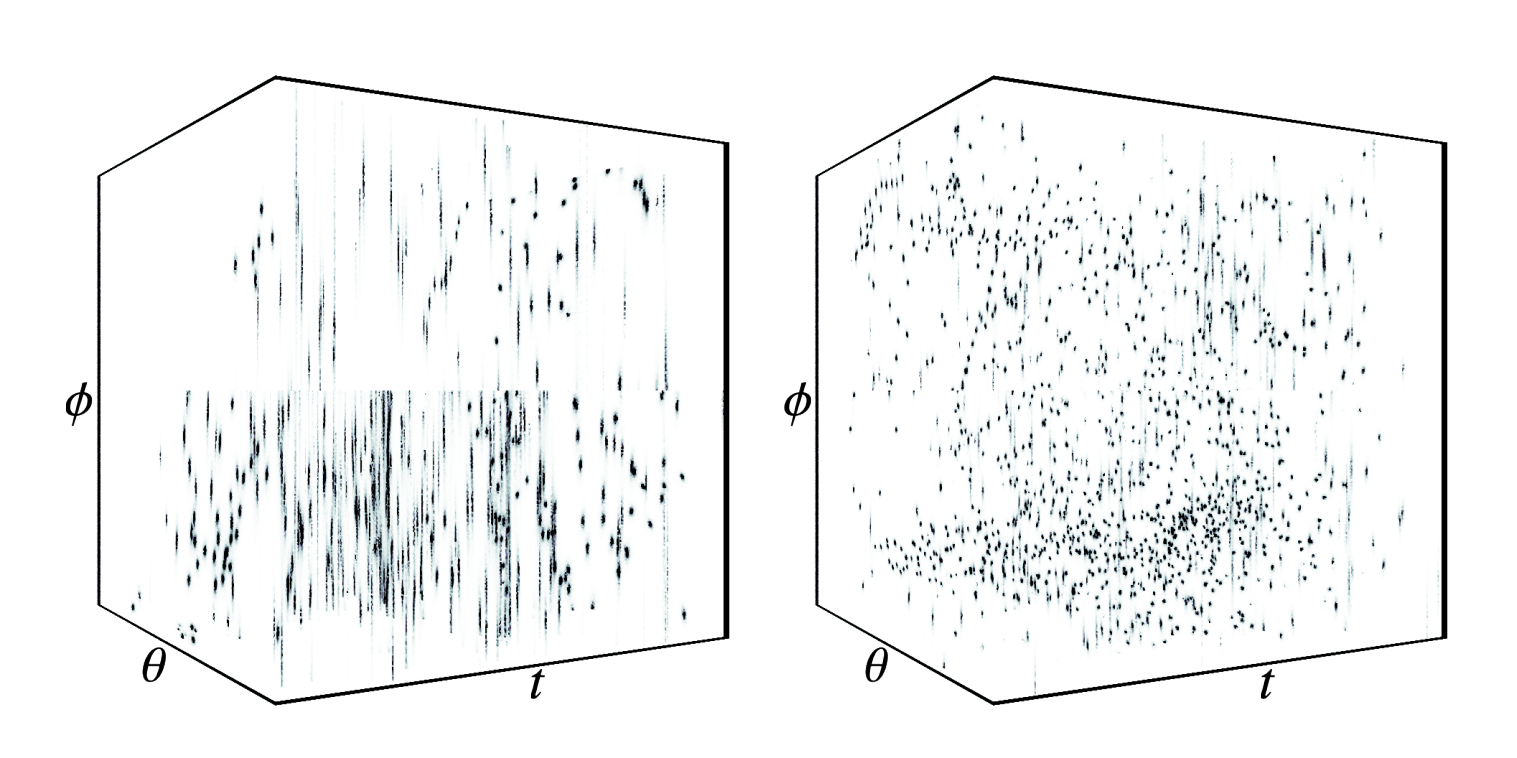}
    \end{tabular}
    
    \caption{\label{fig:boundary_sample}
        \textbf{Boundary sample space:} visualization of the boundary integrand. Note that the integrand is often very sparse and has high frequency, which makes it very hard to sample. Here, \( t \in \mathcal{E} \) is a position on edge, \( (\theta, \phi) \) is a spherical direction. Reference: Figure 7 in \cite{Zhang:2023:Projective}.
    }
\end{figure}

In this section, we will discuss some strategies for efficiently estimating the boundary term of Eq.~\ref{eq:diff_path_int}. This is a term unique to differentiable rendering, or more precisely, to differential spherical integral (\S\ref{sec:theory_spherical}) and differential path integral (\S\ref{sec:theory_psdr}). Since differential path integral generally offers better performance and robustness, we will be focusing on this theory in this section. 

As mentioned in \S\ref{sec:theory_psdr}, PSDR \cite{Zhang:2020:PSDR} rewrites the boundary integral into Eq.~\ref{eq:psdr_rewrite_xw}, and proposes to sample boundary light paths in a multi-directional manner. To sample the boundary segment \( (\ps_0, \pd_0) \), we need to sample a point \( \xb \) on an edge (a 1D manifold), followed by a direction \( \omegabnd \) (a 2D manifold). From the view of sampling, we can rewrite the integral in Eq.~\ref{eq:psdr_rewrite_xw} as a primary-sample-space integral:
\begin{equation}
\label{eq:boundary_primary_sample}
\int_{[0, 1)^3} F(u_1, u_2) \D u_2 \D u_1,
\end{equation}
where \( u_1 \in \mathbb{R} \), \( u_2 \in \mathbb{R}^2 \), and \( F(u_1, u_2) \) is the original integrand changing variable from \( \xb \) and \( \omegabnd \) to \( u_1 \) and \( u_2 \). 

Note that the integrand \( F \) is often sparse and exhibits high frequency \cite{Zhang:2023:Projective}, as demonstrated in Figure~\ref{fig:boundary_sample}, making it very hard to sample. To efficiently sample \( u_1 \) and \( u_2 \), Zhang et al. \cite{Zhang:2020:PSDR} introduce nexthibits-event estimation (NEE) and importance sampling using a pre-processed grid-based structure. In the pre-processing step, a photon mapping (from the light sources) and importon mapping (from the sensor) pass is performed, giving each grid an estimated importance. This spatial importance information is then utilized to sample the edge and complete the light path.

Although the importance sampling technique in PSDR is effective in reducing the variance of estimation, when the scene becomes more complicated, the resolution of the grid has to be impractically high in order to achieve a desirable result. To address this issue, Yan et al.~\cite{Yan:2022:Guiding} propose a kd-tree guiding structure in replacement of the regular grid structure used in PSDR, which improves scalability to more complex scenes. They also apply multiple importance sampling instead of pure emitter sampling for NEE, to better handle complex lighting (e.g., image-based environmental lighting) where emitter sampling is likely to fail due to the emitter being occluded. In addition, an edge sorting step is introduced to form chains of edges that require less subdivision during kd-tree construction, hence improving performance of the guiding step.

\setlength{\resLen}{3.4in}
\begin{figure}[t]
    \centering
    \small
    \addtolength{\tabcolsep}{-3pt}
    \begin{tabular}{c}
         \includegraphics[trim={20 15 0 0},clip, width=\resLen]{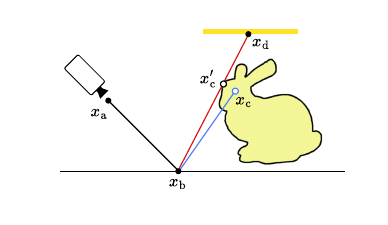}
    \end{tabular}
    
    \caption{\label{fig:boundary_projective}
        \textbf{Projective sampling:} this sampling strategy projects a sampled point at interior \( x_\mathrm{c} \) to a nearby point \( x_\mathrm{c}' \) on the boundary to construct a boundary light path. Reference: Figure 5 in \cite{Zhang:2023:Projective}.
    }
\end{figure}

Based on the idea of differential path integrals, Zhang et al.~\cite{Zhang:2023:Projective} propose another strategy to sample the boundary light paths, known as projective sampling. The core of their method lies in projection operators, tailored to the underlying geometric representation, which map primal samples onto nearby boundaries, ensuring that boundaries near high-contributing regions are sampled more frequently. Figure~\ref{fig:boundary_projective} illustrates this process. By simplifying the local boundary formulation from PSDR and extending it to handle interior regions, it allows effective projection of samples across various geometric representations, including triangle meshes, signed distance functions, and curve-based fibers. They use octrees for their guiding structure, which further optimizes the sampling process. This approach improves the accuracy of gradient estimation by concentrating sampling efforts on areas with significant visibility changes, thereby reducing variance and improving overall optimization efficiency.

\section{Conclusion}
\label{sec:conclusion}

In this survey, we present the current state-of-the-art of physics-based differentiable rendering. We introduce general theories that ensure unbiased estimation of the differential rendering equations, sampling strategies that aim to lower variance, and differentiation techniques that optimize the gradient propagation process. In the current state, the general theories of handling discontinuities have branched into two different routes: explicitly sampling discontinuities or avoiding sampling them by reparameterization. PSDR \cite{Zhang:2020:PSDR} and warped-area reparameterization for differential path integral \cite{Xu:2023:PSDR-WAS} represent the most advanced theories of these two routes. In addition, theories focusing on implicit surfaces have also greatly broadened the range of potential applications for physics-based differentiable rendering. 

Apart from unbiasedness of the estimation, achieving low variance is also another primal goal in rendering. Therefore, studies have proposed specialized estimators for BSDFs and pixel reconstruction filters to efficiently estimate the interior integral. In addition, optimizations in gradient propagation have also proven to be very helpful in improving overall performance. For the boundary integral, custom guiding structures and sampling strategies are introduced. Many of these studies take inspiration from traditional forward rendering studies, and successfully apply them to differentiable rendering. 

The field of physics-based differentiable rendering is still very young and active, and is receiving increasing attention due to its potential in various inverse rendering tasks. Generalizing the theory to more scenarios, proposing better variance reduction techniques, and optimizing the rendering process on system or hardware levels would be promising directions for future studies.

\bibliographystyle{eg-alpha-doi} 
\bibliography{star}
\end{document}